\documentclass[%
reprint,
superscriptaddress,
nofootinbib,
amsmath,amssymb,prl,showpacs
aps
]{revtex4-2}
\usepackage[utf8]{inputenc}
\usepackage[T1]{fontenc}

\usepackage{graphicx}
\usepackage{dcolumn}
\usepackage{bm}
\usepackage{siunitx}
\usepackage{xspace}
\usepackage[dvipsnames]{xcolor}
\usepackage{todonotes}
\usepackage{booktabs}
\usepackage{hyperref}
\usepackage[capitalize]{cleveref}
\usepackage{subcaption}
\usepackage{multirow}
\colorlet{RED}{red}

\makeatletter
\renewcommand*{\@fnsymbol}[1]{\ensuremath{\ifcase#1\or *\or \dagger\or \ddagger\or
   \mathsection\or \mathparagraph\or \|\or **\or \dagger\dagger
   \or \ddagger\ddagger \or \mathparagraph\mathparagraph \else\@ctrerr\fi}}
\makeatother

\providecommand{\ssqthonethree}{\ensuremath{\sin^2\theta_{13}}\xspace}
\providecommand{\ssqthtwothree}{\ensuremath{\sin^2\theta_{23}}\xspace}
\providecommand{\ssqthonetwo}{\ensuremath{\sin^2\theta_{12}}\xspace}
\providecommand{\deltacp}{\ensuremath{\delta_{\scriptscriptstyle\mathrm{CP}}}\xspace}
\providecommand{\jarlcp}{\ensuremath{J_{\scriptscriptstyle\mathrm{CP}}}\xspace}

\providecommand{\dmsqtwothree}{\ensuremath{\Delta{}m^2_{32}}\xspace}
\providecommand{\dmsqonetwo}{\ensuremath{\Delta{}m^2_{21}}\xspace}

\providecommand{\nubar}{\ensuremath{\bar{\nu}}\xspace}
\providecommand{\nue}{\ensuremath{\nu_{e}}\xspace}
\providecommand{\numu}{\ensuremath{\nu_{\mu}}\xspace}
\providecommand{\nueb}{\ensuremath{\nubar_{e}}\xspace}
\providecommand{\numub}{\ensuremath{\nubar_{\mu}}\xspace}

\begin{document}

\title{Results from the T2K experiment on neutrino mixing including a new far detector $\mu$-like sample}


\newcommand{\INSTHD}{\affiliation{University Autonoma Madrid, Department of Theoretical Physics, 28049 Madrid, Spain}}
\newcommand{\INSTFE}{\affiliation{Boston University, Department of Physics, Boston, Massachusetts, U.S.A.}}
\newcommand{\INSTD}{\affiliation{University of British Columbia, Department of Physics and Astronomy, Vancouver, British Columbia, Canada}}
\newcommand{\INSTGA}{\affiliation{University of California, Irvine, Department of Physics and Astronomy, Irvine, California, U.S.A.}}
\newcommand{\INSTI}{\affiliation{IRFU, CEA, Universit\'e Paris-Saclay, F-91191 Gif-sur-Yvette, France}}
\newcommand{\INSTGB}{\affiliation{University of Colorado at Boulder, Department of Physics, Boulder, Colorado, U.S.A.}}
\newcommand{\INSTFH}{\affiliation{Duke University, Department of Physics, Durham, North Carolina, U.S.A.}}
\newcommand{\INSTJA}{\affiliation{E\"{o}tv\"{o}s Lor\'{a}nd University, Department of Atomic Physics, Budapest, Hungary}}
\newcommand{\INSTEF}{\affiliation{ETH Zurich, Institute for Particle Physics and Astrophysics, Zurich, Switzerland}}
\newcommand{\INSTIG}{\affiliation{VNU University of Science, Vietnam National University, Hanoi, Vietnam}}
\newcommand{\INSTIE}{\affiliation{CERN European Organization for Nuclear Research, CH-1211 Gen\'eve 23, Switzerland}}
\newcommand{\INSTEG}{\affiliation{University of Geneva, Section de Physique, DPNC, Geneva, Switzerland}}
\newcommand{\INSTHJ}{\affiliation{University of Glasgow, School of Physics and Astronomy, Glasgow, United Kingdom}}
\newcommand{\INSTJG}{\affiliation{Ghent University, Department of Physics and Astronomy, Proeftuinstraat 86, B-9000 Gent, Belgium}}
\newcommand{\INSTDG}{\affiliation{H. Niewodniczanski Institute of Nuclear Physics PAN, Cracow, Poland}}
\newcommand{\INSTCB}{\affiliation{High Energy Accelerator Research Organization (KEK), Tsukuba, Ibaraki, Japan}}
\newcommand{\INSTIB}{\affiliation{University of Houston, Department of Physics, Houston, Texas, U.S.A.}}
\newcommand{\INSTED}{\affiliation{Institut de Fisica d'Altes Energies (IFAE) - The Barcelona Institute of Science and Technology, Campus UAB, Bellaterra (Barcelona) Spain}}
\newcommand{\INSTJC}{\affiliation{Institut f\"ur Physik, Johannes Gutenberg-Universit\"at Mainz, Staudingerweg 7, 55128 Mainz, Germany}}
\newcommand{\INSTHH}{\affiliation{Institute For Interdisciplinary Research in Science and Education (IFIRSE), ICISE, Quy Nhon, Vietnam}}
\newcommand{\INSTEI}{\affiliation{Imperial College London, Department of Physics, London, United Kingdom}}
\newcommand{\INSTGF}{\affiliation{INFN Sezione di Bari and Universit\`a e Politecnico di Bari, Dipartimento Interuniversitario di Fisica, Bari, Italy}}
\newcommand{\INSTBE}{\affiliation{INFN Sezione di Napoli and Universit\`a di Napoli, Dipartimento di Fisica, Napoli, Italy}}
\newcommand{\INSTBF}{\affiliation{INFN Sezione di Padova and Universit\`a di Padova, Dipartimento di Fisica, Padova, Italy}}
\newcommand{\INSTBD}{\affiliation{INFN Sezione di Roma and Universit\`a di Roma ``La Sapienza'', Roma, Italy}}
\newcommand{\INSTEB}{\affiliation{Institute for Nuclear Research of the Russian Academy of Sciences, Moscow, Russia}}
\newcommand{\INSTHI}{\affiliation{International Centre of Physics, Institute of Physics (IOP), Vietnam Academy of Science and Technology (VAST), 10 Dao Tan, Ba Dinh, Hanoi, Vietnam}}
\newcommand{\INSTJD}{\affiliation{ILANCE, CNRS – University of Tokyo International Research Laboratory, Kashiwa, Chiba 277-8582, Japan}}
\newcommand{\INSTHA}{\affiliation{Kavli Institute for the Physics and Mathematics of the Universe (WPI), The University of Tokyo Institutes for Advanced Study, University of Tokyo, Kashiwa, Chiba, Japan}}
\newcommand{\INSTID}{\affiliation{Keio University, Department of Physics, Kanagawa, Japan}}
\newcommand{\INSTIF}{\affiliation{King's College London, Department of Physics, Strand, London WC2R 2LS, United Kingdom}}
\newcommand{\INSTCC}{\affiliation{Kobe University, Kobe, Japan}}
\newcommand{\INSTCD}{\affiliation{Kyoto University, Department of Physics, Kyoto, Japan}}
\newcommand{\INSTEJ}{\affiliation{Lancaster University, Physics Department, Lancaster, United Kingdom}}
\newcommand{\INSTII}{\affiliation{Lawrence Berkeley National Laboratory, Berkeley, California, U.S.A.}}
\newcommand{\INSTBA}{\affiliation{Ecole Polytechnique, IN2P3-CNRS, Laboratoire Leprince-Ringuet, Palaiseau, France}}
\newcommand{\INSTFC}{\affiliation{University of Liverpool, Department of Physics, Liverpool, United Kingdom}}
\newcommand{\INSTFI}{\affiliation{Louisiana State University, Department of Physics and Astronomy, Baton Rouge, Louisiana, U.S.A.}}
\newcommand{\INSTIH}{\affiliation{Joint Institute for Nuclear Research, Dubna, Moscow Region, Russia}}
\newcommand{\INSTHB}{\affiliation{Michigan State University, Department of Physics and Astronomy,  East Lansing, Michigan, U.S.A.}}
\newcommand{\INSTCE}{\affiliation{Miyagi University of Education, Department of Physics, Sendai, Japan}}
\newcommand{\INSTDF}{\affiliation{National Centre for Nuclear Research, Warsaw, Poland}}
\newcommand{\INSTFJ}{\affiliation{State University of New York at Stony Brook, Department of Physics and Astronomy, Stony Brook, New York, U.S.A.}}
\newcommand{\INSTEH}{\affiliation{STFC, Rutherford Appleton Laboratory, Harwell Oxford,  and  Daresbury Laboratory, Warrington, United Kingdom}}
\newcommand{\INSTGJ}{\affiliation{Okayama University, Department of Physics, Okayama, Japan}}
\newcommand{\INSTCF}{\affiliation{Osaka Metropolitan University, Department of Physics, Osaka, Japan}}
\newcommand{\INSTGG}{\affiliation{Oxford University, Department of Physics, Oxford, United Kingdom}}
\newcommand{\INSTIC}{\affiliation{University of Pennsylvania, Department of Physics and Astronomy,  Philadelphia, Pennsylvania, U.S.A.}}
\newcommand{\INSTGC}{\affiliation{University of Pittsburgh, Department of Physics and Astronomy, Pittsburgh, Pennsylvania, U.S.A.}}
\newcommand{\INSTGD}{\affiliation{University of Rochester, Department of Physics and Astronomy, Rochester, New York, U.S.A.}}
\newcommand{\INSTHC}{\affiliation{Royal Holloway University of London, Department of Physics, Egham, Surrey, United Kingdom}}
\newcommand{\INSTBC}{\affiliation{RWTH Aachen University, III. Physikalisches Institut, Aachen, Germany}}
\newcommand{\INSTJF}{\affiliation{School of Physics and Astronomy, University of Minnesota, Minneapolis, Minnesota, U.S.A.}}
\newcommand{\INSTJB}{\affiliation{Departamento de F\'isica At\'omica, Molecular y Nuclear, Universidad de Sevilla, 41080 Sevilla, Spain}}
\newcommand{\INSTFB}{\affiliation{University of Sheffield, School of Mathematical and Physical Sciences, Sheffield, United Kingdom}}
\newcommand{\INSTDI}{\affiliation{University of Silesia, Institute of Physics, Katowice, Poland}}
\newcommand{\INSTIA}{\affiliation{SLAC National Accelerator Laboratory, Stanford University, Menlo Park, California, U.S.A.}}
\newcommand{\INSTBB}{\affiliation{Sorbonne Universit\'e, CNRS/IN2P3, Laboratoire de Physique Nucl\'eaire et de Hautes Energies (LPNHE), Paris, France}}
\newcommand{\INSTJE}{\affiliation{South Dakota School of Mines and Technology, 501 East Saint Joseph Street, Rapid City, SD 57701, United States}}
\newcommand{\INSTCH}{\affiliation{University of Tokyo, Department of Physics, Tokyo, Japan}}
\newcommand{\INSTBJ}{\affiliation{University of Tokyo, Institute for Cosmic Ray Research, Kamioka Observatory, Kamioka, Japan}}
\newcommand{\INSTCG}{\affiliation{University of Tokyo, Institute for Cosmic Ray Research, Research Center for Cosmic Neutrinos, Kashiwa, Japan}}
\newcommand{\INSTHF}{\affiliation{Institute of Science Tokyo, Department of Physics, Tokyo}}
\newcommand{\INSTGI}{\affiliation{Tokyo Metropolitan University, Department of Physics, Tokyo, Japan}}
\newcommand{\INSTHG}{\affiliation{Tokyo University of Science, Faculty of Science and Technology, Department of Physics, Noda, Chiba, Japan}}
\newcommand{\INSTB}{\affiliation{TRIUMF, Vancouver, British Columbia, Canada}}
\newcommand{\INSTJH}{\affiliation{University of Toyama, Department of Physics, Toyama, Japan}}
\newcommand{\INSTDJ}{\affiliation{University of Warsaw, Faculty of Physics, Warsaw, Poland}}
\newcommand{\INSTDH}{\affiliation{Warsaw University of Technology, Institute of Radioelectronics and Multimedia Technology, Warsaw, Poland}}
\newcommand{\INSTIJ}{\affiliation{Tohoku University, Faculty of Science, Department of Physics, Miyagi, Japan}}
\newcommand{\INSTFD}{\affiliation{University of Warwick, Department of Physics, Coventry, United Kingdom}}
\newcommand{\INSTEA}{\affiliation{Wroclaw University, Faculty of Physics and Astronomy, Wroclaw, Poland}}
\newcommand{\INSTHE}{\affiliation{Yokohama National University, Department of Physics, Yokohama, Japan}}
\newcommand{\INSTH}{\affiliation{York University, Department of Physics and Astronomy, Toronto, Ontario, Canada}}

\INSTHD
\INSTFE
\INSTD
\INSTGA
\INSTI
\INSTGB
\INSTFH
\INSTJA
\INSTEF
\INSTIG
\INSTIE
\INSTEG
\INSTHJ
\INSTJG
\INSTDG
\INSTCB
\INSTIB
\INSTED
\INSTJC
\INSTHH
\INSTEI
\INSTGF
\INSTBE
\INSTBF
\INSTBD
\INSTEB
\INSTHI
\INSTJD
\INSTHA
\INSTID
\INSTIF
\INSTCC
\INSTCD
\INSTEJ
\INSTII
\INSTBA
\INSTFC
\INSTFI
\INSTIH
\INSTHB
\INSTCE
\INSTDF
\INSTFJ
\INSTEH
\INSTGJ
\INSTCF
\INSTGG
\INSTIC
\INSTGC
\INSTGD
\INSTHC
\INSTBC
\INSTJF
\INSTJB
\INSTFB
\INSTDI
\INSTIA
\INSTBB
\INSTJE
\INSTCH
\INSTBJ
\INSTCG
\INSTHF
\INSTGI
\INSTHG
\INSTB
\INSTJH
\INSTDJ
\INSTDH
\INSTIJ
\INSTFD
\INSTEA
\INSTHE
\INSTH

\author{K.\,Abe}\INSTBJ
\author{S.\,Abe}\INSTCH
\author{R.\,Akutsu}\INSTCB
\author{H.\,Alarakia-Charles}\INSTEJ
\author{Y.I.\,Alj Hakim}\INSTFB
\author{S.\,Alonso Monsalve}\INSTEF
\author{L.\,Anthony}\INSTEI
\author{S.\,Aoki}\INSTCC
\author{K.A.\,Apte}\INSTEI
\author{T.\,Arai}\INSTCH
% ...

\collaboration{The T2K Collaboration}\noaffiliation

\date{\today}

\begin{abstract}
We have made improved measurements of three-flavor neutrino mixing with 19.7(16.3)$\times 10^{20}$ protons on target in (anti-)neutrino-enhanced beam modes. A new sample of muon-neutrino events with tagged pions has been added at the far detector, as well as new proton and photon-tagged samples at the near detector. Significant improvements have been made to the flux and neutrino interaction modeling. T2K data continue to prefer the normal mass ordering and upper octant of $\sin^2\theta_{23}$ with a near-maximal value of the charge-parity violating phase with best-fit values in the normal ordering of \deltacp$=-2.18\substack{+1.22 \\ -0.47}$, \ssqthtwothree$=0.559\substack{+0.018 \\ -0.078}$ and \dmsqtwothree$=(+2.506\substack{+0.039 \\ -0.052})\times 10^{-3}$ eV$^{2}$.
\end{abstract}

\maketitle

\indent{{\bf {\em Introduction---}}} The T2K experiment~\cite{ABE2011106} measures three-flavor neutrino mixing parameters by observing the disappearance of \numu (\numub) and the appearance of \nue (\nueb) over a distance of 295 km in a narrow-band, predominantly \numu or \numub beam, which peaks at an energy of 0.6 GeV. Data corresponding to 3.6$\times$10$^{21}$ protons on target (POT) were analyzed with major improvements to the neutrino flux and interaction modeling compared to previous studies \cite{T2K:2023smv}. New photon- and proton-tagged near detector (ND) samples, as well as a new, charged current neutrino-enhanced muon-neutrino sample with tagged pions ($\nu_\mu$CC$1\pi^+$-like) in the far detector, are included in the analysis for the first time. This letter reports T2K's latest measurements of neutrino mixing parameters after these improvements, which mark a significant step forward in the robustness of our analysis, particularly due to improvements in modeling and constraining neutrino interaction uncertainties, and supersede previous results produced with the same dataset.
 Comparable measurements have also been reported by NOvA~\cite{NOvA:2021nfi}, and future experiments such as Hyper-Kamiokande~\cite{Hyper-Kamiokande:2018ofw} and DUNE~\cite{DUNE:2020ypp} aim to further improve the precision and discovery potential in neutrino oscillation physics.

\indent{{\bf {\em The T2K experiment---}}} 
Protons of 30 GeV energy, accelerated by the J-PARC main ring, collide and interact with a graphite target, producing pions and kaons that are focused by a system of three magnetic horns and decay inside a 96 m long tunnel. Depending on the direction of the current flowing in the horns, either positively or negatively charged hadrons are focused, producing a neutrino- ($\nu$-mode) or antineutrino-enhanced ($\bar{\nu}$-mode) beam.

Two near detectors are located 280~m away from the graphite target --- one on-axis (INGRID)~\cite{Abe:2011xv}, and the other 2.5$^\circ$ off-axis (ND280)~\cite{ABE2011106} with respect to the beam direction. They sample the unoscillated beam by monitoring its direction, intensity and flavor content as well as constraining uncertainties in the neutrino interaction model. The far detector, Super-Kamiokande (SK)~\cite{FUKUDA2003418}, is a 50~kton water-Cherenkov detector located beneath a 1~km rock overburden within the Kamioka mine in Japan. It measures the oscillated neutrino flux 295 km away from its production point at $2.5^\circ$ off-axis.

The analysis reported in this letter uses all data collected by T2K from January 2010 to February 2020, corresponding to an SK exposure of $19.7\times10^{20}$~POT in $\nu$-mode and $16.3\times10^{20}$~POT in $\bar{\nu}$-mode. The same number of POT was used in a previous analysis~\cite{T2K:2023smv}.

\indent{{\bf {\em Neutrino flux prediction---}}} The neutrino flux Monte Carlo (MC) simulation uses the FLUKA 2011.2~\cite{10.3389/fphy.2021.788253, BATTISTONI201510} interaction model for proton interactions inside the target and GEANT3~\cite{GEANT3} for interactions outside the target. Hadron production in the target is tuned to external datasets~\cite{Abgrall:2015hmv, Abgrall:2016jif, NA61SHINE:2018rhe}. The flux depends on the beamline conditions and the measured proton beam profile. The prediction accounts for the beam conditions during each operating period.

The resulting flux model~\cite{Abe:2012av} is used to estimate the unoscillated $\nu$-mode and $\bar{\nu}$-mode fluxes at all detectors, for all contributing neutrino flavors, as well as their correlated uncertainties. INGRID is used to monitor the beam direction and validate the neutrino flux simulation. The uncertainty on the measured beam direction is included in the flux uncertainties used in the oscillation parameter measurement.

This analysis includes new constraints from the NA61/SHINE replica target datasets~\cite{NA61SHINE:2018rhe}, including high-statistics measurements of $\pi^{\pm}$ production as well as, for the first time, $K^{\pm}$ and proton production measurements. Among other improvements these constraints reduce the \numu flux uncertainty to below 4\% for energies up to 7~GeV.

\indent{{\bf {\em Neutrino interaction modeling---}}} The nominal model prediction was generated with the neutrino event generator NEUT 5.4.0~\cite{Hayato:2021heg}. In this analysis, important improvements to the estimated neutrino interaction uncertainties were made, increasing the number of free parameters by 26 compared with the previous analysis~\cite{T2K:2023smv}, for a total of 75.

\paragraph{Pionless:} New degrees of freedom were included to cover theoretical uncertainties in nuclear effects for Charged-Current (CC) Quasi-Elastic (CCQE), multi-nucleon (2p2h) and hard-scattering processes.

\emph{Ad hoc} uncertainties in the low four-momentum transfer ($Q^{2}$) response for CCQE interactions were replaced with physics motivated parameters that vary the impact of Pauli blocking and the effect of nuclear transparency on the inclusive cross section. An effective parameterization of the latter was derived by comparison to the NuWro~\cite{NuWro:GOLAN2012499, Juszczak:2009qa} implementation of the optical potential~\cite{ankowski:PRD91-2014}. New uncertainties on the nuclear ground state for CCQE events, which parameterize the shell structure of the missing momentum ($p_\text{miss}$) and missing energy ($E_\text{miss}$) response in the initial state Spectral Function (SF), were included~\cite{Stowell:2016jfr, Chakrani:2023htw}. The Benhar SF~\cite{benharsf} incorporates multi-nucleon knockout in the $E_\text{miss}>100$~MeV, $p_\text{miss}>300$~MeV region. These events account for 5\% of all CCQE interactions; a 200\% normalization uncertainty was applied to cover differences between NEUT and NuWro predictions. Following Ref.~\cite{bodekcai:EPJC2019}, a three-momentum-transfer-dependent freedom on the Nucleon Removal Energy (NRE) was added~\cite{dolan2023electronnucleus}. Additional freedom for 2p2h interactions was included by separating some of the existing uncertainties by the struck nucleon pair (proton--neutron vs. proton--proton/neutron--neutron).

\paragraph{Single pion production (SPP):} Uncertainties on the free parameters in the NEUT implementation~\cite{T2K:2023smv} of the Rein--Sehgal SPP model~\cite{Rein:1980wg} were motivated by fitting~\cite{Stowell:2016jfr} to hydrogen- and deuterium-target data~\cite{Wilkinson:2014yfa, Rodrigues:2016xjj} from ANL~\cite{anl1:PhysRevD.23.569,anl2:PhysRevD.25.1161} and BNL~\cite{bnl1:PhysRevD.23.2495,bnl2:PhysRevD.34.2554,bnl3:Furuno:2003ng}, and checking coverage against hydrocarbon-target data~\cite{MiniBooNE:2010eis, MiniBooNE:2009dxl, MiniBooNE:2010cxl, MINERvA:2014ogb, MINERvA:2015slz, MINERvA:2017okh, MINERvA:2016sfc}. Variations of these parameters exhibit a minimal change to both the shape of the final-state pion kinematics, which have previously been observed to be poorly predicted at both ND280 and SK~\cite{T2K:2023smv}, and to the relative rate of charged and neutral pion production. To cover measurements of pion kinematic spectra in T2K \cite{T2K:2019yqu} and MINER$\nu$A \cite{MINERvA:2019kfr}, new parameters were developed to vary the resonance decay kinematics and the total neutral pion production rate~\cite{Stowell:2016jfr}. Previously, NRE effects were only included in NEUT for pionless hard-scatter processes. Here, an approximation of this effect, and an associated uncertainty, was motivated by comparing NEUT predictions to those from NuWro for $0 < \text{NRE}_{\mathrm{SPP}} < 50$~MeV.

\paragraph{Multiple pions:} New uncertainties were developed for multiple pion production in the low hadronic mass region, $1.3 < W < 2.0~\mathrm{GeV}$. Ref.~\cite{Bodek:2021} motivated separate uncertainties on the axial and vector parts of the nucleon form factors, replacing an effective 100\%, $Q^2$-dependent, uncertainty on the previous Bodek-Yang correction~\cite{Bodek:2003wc}. Additionally, a new uncertainty covering the difference between the NEUT and AGKY models~\cite{Yang:2009zx} was included on the shape of the two-dimensional pion multiplicity and invariant hadronic mass distributions.

\paragraph{Final-state interactions:} A new 16\% uncertainty on the cross section for pion charge-exchange reactions for $p_\pi > 500$~MeV was included~\cite{PinzonGuerra:2018rju}. As the new ND selections are sensitive to proton kinematics, an uncertainty was added that varies the fraction of events that include intranuclear nucleon rescattering while keeping the leptonic observables unmodified. The size of this uncertainty (30\%) was motivated by a recent analysis of nuclear transparency data~\cite{niewczas:PhysRevC.100.015505}.

\indent{{\bf {\em Near detector analysis---}}} The neutrino flux and interaction models were constrained by fitting the unoscillated CC event spectra at the ND280. Three Time Projection Chambers (TPC)~\cite{T2KND280TPC:2010nnd} with two Fine-Grained Detectors (FGD)~\cite{T2KND280FGD:2012umz} sandwiched between them track particles from interactions in the FGD target mass. Electromagnetic calorimeters surrounding the TPCs and FGDs, as well as the TPCs, are used to tag photons.

Selected CC events were separated into different samples according to the FGD in which the interaction occurred, the beam mode, the muon charge and the reconstructed final-state particle multiplicities. Negatively-charged muon candidates selected in $\nu$-mode were divided into five samples per FGD: 1) at least one photon (CC-photon); 2) no photons, pions or protons (CC0$\pi$0p); 3) no photons or pions and at least one proton (CC0$\pi$Np); 4) no photons and one positively-charged pion (CC1$\pi^{+}$); 5) and all other CC events (CC-Other). Positively and negatively charged muon candidates in $\bar{\nu}$-mode are each divided into three samples per FGD: 1) no pions (CC0$\pi$); 2) one pion with opposite charge to the muon (CC1$\pi$); 3) and all other CC events (CC-Other). Positively charged muon candidates in $\nu$-mode were not separately selected as the predicted contamination was 4\%, compared to 30\% negatively charged muon contamination in $\bar{\nu}$-mode~\cite{PhysRevD.96.092006}.

Newly added proton-tagged samples offer enhanced abilities to constrain the $Q^2$-dependence of several uncertainties with data. Since ND280 achieves reliable proton tracking efficiency for momenta $\gtrsim$450 MeV, samples with (without) protons generally correspond to higher (lower) average $Q^2$. The new photon-tagged sample helps to constrain $\pi^0$ production uncertainties, and increases the purity of other samples.

\begin{figure}[htbp]
  \centering
  \includegraphics[width=0.4\textwidth]{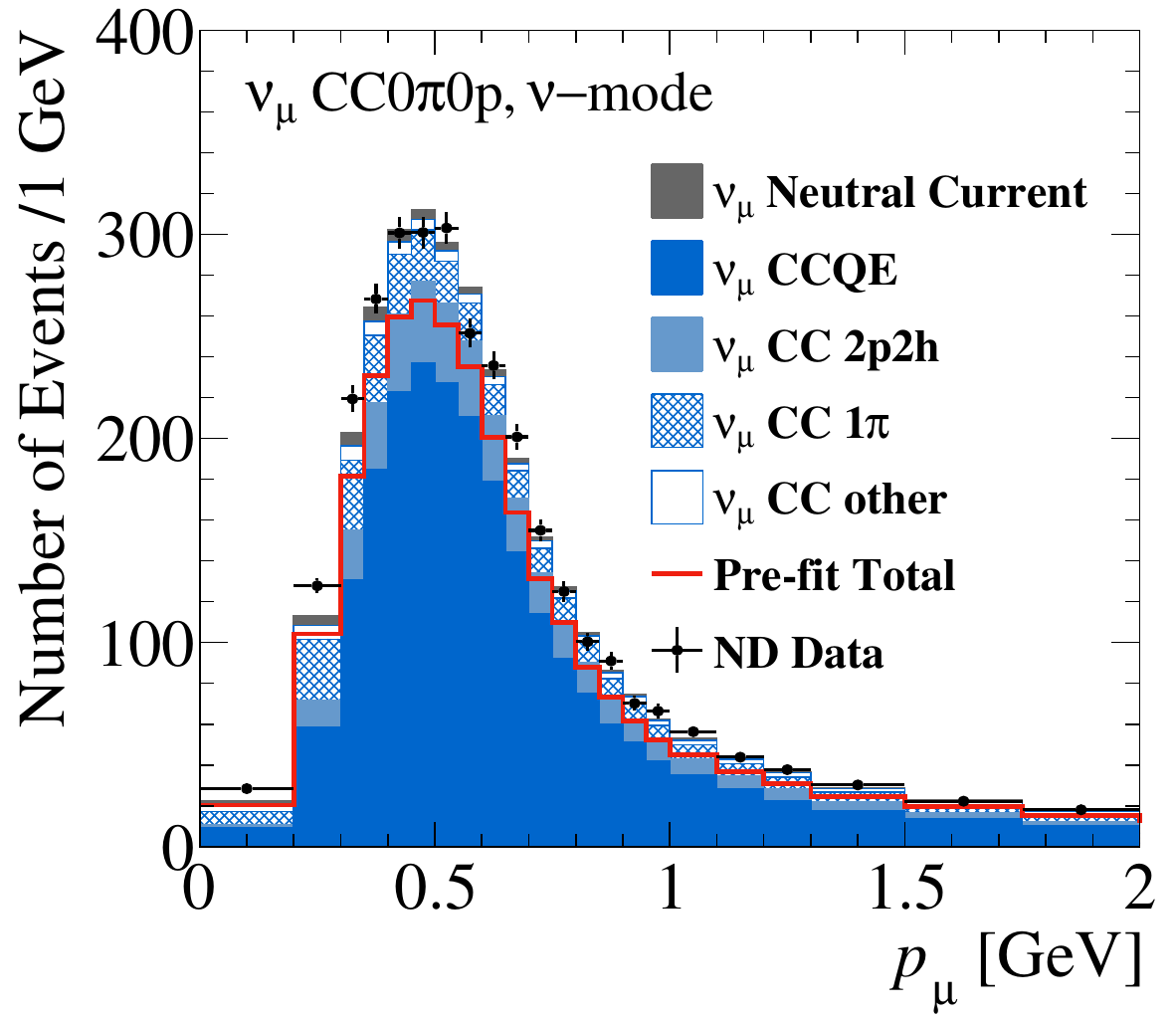}
  \caption{Data and model predictions before and after the ND280 fit for the $\nu$-mode FGD1 CC0$\pi$0p sample binned in $p_{\mu}$. The fit includes $p_{\mu} \leq 30$ GeV for all ND280 samples, but the range is truncated here for readability. The postfit prediction is broken down by interaction channel. The predicted event rate from $\bar{\nu}$ contamination in $\nu$-mode is neglected in the figure as it only contributes at the sub-percent level.}
\label{Fig::CC0pi0p}
\end{figure}

Each sample was binned according to the muon candidate's momentum, $p_{\mu}$, and the cosine of the angle between the muon momentum vector and the beam direction, $\cos\theta_{\mu}$. A MC prediction was made using the flux, cross-section and ND280 detector models, and an extended binned likelihood~\cite{Conway:2011in} fit to data was performed. \autoref{Fig::CC0pi0p} shows the data, prefit and postfit MC distributions for the FGD1 CC0$\pi$0p sample as a function of $p_{\mu}$. The total $p$-value is 0.575, indicating agreement between the data fit result and the input models. The uncertainty on the SK predicted event rates from cross-section and flux systematics was reduced from approximately 10--15\% to 3--4\% for each sample by using the ND280 constraint.

\indent{{\bf {\em SK event selection---}}}Previous T2K analyses used five SK samples: 1 ring $\mu$- and $e$-like in both $\nu$- and $\bar{\nu}$-modes and 1 ring $e$-like with 1 decay electron in $\nu$-mode~\cite{T2K:2023smv} (tagging a pion below Cherenkov threshold). This analysis introduces a $\nu_\mu$CC1$\pi^+$-like sample in $\nu$-mode, which tags pions through two topologies: (1) one ring each from a muon and a charged pion with one or two decay electrons (from the pion and muon decay); and (2) one $\mu$-like ring (where the charged pion is below Cherenkov threshold) with two decay electrons (from the decay of the muon and the pion). Standard pre-selection criteria common to all SK samples~\cite{T2K:2023smv} were applied in addition to the particle-identification (PID) requirements for the candidate $\mu$-like, $\pi$-like, and decay electron rings. These PID requirements reduce the number of selected background events with primary electrons or neutral pions, and those produced by neutral-current pion production processes.

\begin{figure}[htbp]
  \centering
  \includegraphics[width=0.4\textwidth]{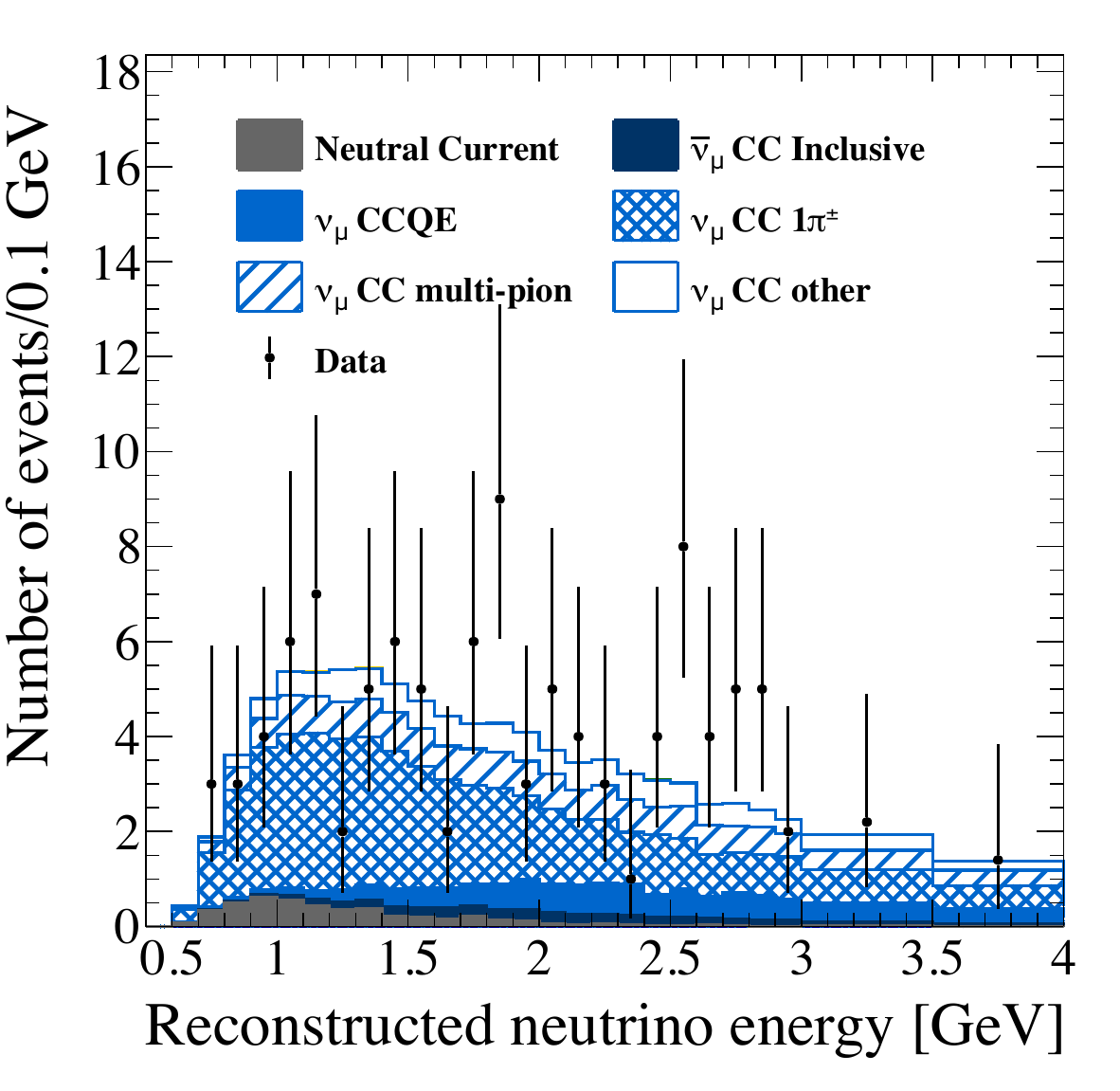}
  \caption{The $E^\text{rec}_{\nu}$ distribution for the $\nu_{\mu}$CC1$\pi^+$ $\nu$-mode SK sample shown for data and MC.  The fit includes $\mu$-like ($e$-like) events with $E^\text{rec}_{\nu} \leq 30$ (1.25) GeV, but the range is truncated here for readability. The reference oscillation parameter values are described in the main text. The ND280 constraint has been applied to the MC prediction.}
  \label{fig:FDMRSample}
\end{figure}

The addition of the $\nu_\mu$CC1$\pi^+$-like sample increases the total number of selected $\mu$-like events by 42.5\%, although much of the increase is above the oscillation maximum, and is affected by different systematics to the dominant 1 ring $\mu$-like samples. At a reference set of oscillation parameter values~\cite{PDG2021} ($\sin^2\theta_{23} = 0.561 $, $\sin^2\theta_{13} = 0.022$, $\sin^2\theta_{12} = 0.307$, $\Delta{m}^{2}_{32} = 2.49\times10^{-3}$~eV$^2$, $\Delta{m}^2_{21} = 7.53\times10^{-5}$~eV$^2$, $\deltacp = -1.601$ and normal ordering), 53.5 (116.6) signal (total) events were predicted in this sample. A total of 134 data events passed the selection criteria with a total systematic uncertainty of 4.3\%. The reconstructed neutrino energy for this sample, using only the reconstructed muon information, is
\begin{equation*}
    E^\text{rec}_{\nu} = \frac{2m_{p}E_{\mu}+ m_{\Delta^{++}}^{2} - m_{p}^{2} - m_{\mu}^{2} }{2\left(m_{p} -E_{\mu} + \left|  \mathbf{p}_{\mu}\right| \cos\theta_{\mu}\right)}, 
\end{equation*}
where $m_{p}$, $m_{\mu}$, $m_{\Delta^{++}}$ are the proton, muon, and $\Delta$ baryon rest masses; and $E_{\mu}$, $\mathbf{p}_{\mu}$, $\theta_{\mu}$ are the muon-candidate reconstructed energy, three-momentum, and angle with respect to the neutrino beam. Predicted and observed $E^\text{rec}_{\nu}$ distributions for this sample are presented in \autoref{fig:FDMRSample}. Since events do not populate the region of the oscillation dip, we do not expect a significant increase in sensitivity to oscillation parameters. However, this sample provides improved control of background contributions and allows for valuable cross-checks of the cross-section model.

\indent{{\bf {\em Oscillation Analysis---}}} The PDG parameterization of the PMNS matrix~\cite{PDG2021} was used in the analysis. Matter effects were included with an average Earth crust density of 2.6~g/cm$^{3}$. The peak neutrino energy and baseline used by T2K provide sensitivity to the PMNS parameters \ssqthonethree, \ssqthtwothree, \deltacp, and the magnitude and sign of the mass-splitting term \dmsqtwothree.

Measurements were carried out with and without a Reactor Constraint (RC). The RC was included as a Gaussian prior on \ssqthonethree~$\sim (2.20 \pm 0.07) \times 10^{-2}$, the PDG world-average~\cite{DoubleChooz:2014kuw,RENO:2018dro,DayaBay:2018yms,PDG2021}. Similarly, for \dmsqonetwo and \ssqthonetwo priors of $(7.53 \pm 0.18) \times 10^{-5}$ \si{eV^{2}} and $(0.307 \pm 0.013)$ were used~\cite{PDG2021}. Flat priors were used for the other oscillation parameters. The impact of the choice of prior on \deltacp was investigated and does not alter the conclusions of the analysis~\cite{T2K:2025stu}.

Two different statistical approaches were used to extract constraints on the oscillation parameters of interest from a likelihood function, the form of which remained unchanged from Ref.~\cite{T2K:2023smv}. One used a Markov Chain Monte Carlo (MCMC) method~\cite{metropolis, hastings}, to simultaneously fit the data from ND280 and SK, producing posterior distributions from which credible intervals were constructed. The other performed a piece-wise fit and marginalized over the parameters associated with the propagated ND280 constraint to provide frequentist confidence intervals, both with the constant $\Delta\chi^{2}$ and Feldman-Cousins (FC)~\cite{Feldman:1997qc} methods.

\autoref{table:GOF_wRC} presents the posterior predictive p-values~\cite{gelman1996posterior, Gelman_Example, Gelman_Understanding} for all FD samples. After accounting for the `Look Elsewhere Effect' using the Bonferroni correction~\cite{Bayer_2020}, all samples pass the 5\% threshold. This suggests that the model provides a plausible description of the data within the considered parameter space.
\begin{table}[htbp]
\centering
\begin{tabular}{ c | c | c }
\hline \hline
\multicolumn{2}{c}{Sample} & $p$-value \\
\hline
\multirow{2}{*}{1R$\mu$-like} & $\nu$-mode & 0.35 \\
 & $\bar{\nu}$-mode & 0.84 \\
\hline
$\nu_{\mu}$CC1$\pi^+$-like & $\nu$-mode & 0.96 \\
\hline
\multirow{2}{*}{1R$e$-like} & $\nu$-mode & 0.13 \\
 & $\bar{\nu}$-mode & 0.63 \\
\hline
1R$e$-like 1d$e$ & $\nu$-mode & 0.89 \\
\hline
\multicolumn{2}{c|}{Total} & 0.86 \\
\hline \hline
\end{tabular}
\caption{Posterior predictive $p$-values for every FD sample, from the fit including the RC.}
\label{table:GOF_wRC}
\end{table}

\indent{{\bf {\em Simulated Data Studies (SDS)---}}} These study the impact of alternative interaction models and data-driven modifications of predictions at ND280 and SK, to check the completeness of our systematic uncertainty model, and are essential tests of the robustness of our result. Changes to \ssqthtwothree and \dmsqtwothree were deemed significant if the bias to the center of a 2$\sigma$ confidence interval was greater than 50\% of the 1$\sigma$ Asimov interval width, or if the size of a one-dimensional, 2$\sigma$, FC-corrected interval changed by more than 10\%. If a significant bias was found, additional smearing was applied to the relevant intervals. For \deltacp, biases to interval boundaries which change the inclusion or exclusion of a physically interesting point in parameter space, \emph{e.g.} $\deltacp = [-\pi,0,\pi]$, are reported as part of the result; but due to the non-Gaussian nature of the \deltacp likelihood additional smearing was not applied to reported intervals.

\begin{table}[htbp]
\centering
\begin{tabular}{c|cccc}
\hline\hline
SDS &  & \ssqthtwothree &\dmsqtwothree & \deltacp \\
\hline
\multirow{2}{*}{HF CRPA}       & Bias & $-25.1$\% & $84.9$\% & $-11.2$\% \\
                               & Size  & $2.0$\%   & $-5.4$\% & $1.0$\%   \\
\hline
\multirow{2}{*}{Martini 1$\pi$} & Bias & $-3.2$\% & $-18.5$\% & $-1.7$\% \\
                                & Size  & $-0.2$\% & $-1.0$\%  & $2.0$\%  \\
\hline
\multirow{2}{*}{Non-CCQE}       & Bias & $10.4$\% & $-76.3$\% & $-0.5$\% \\
                                & Size  & $3.0$\% & $-1.0$\% & $-3.0\%$ \\
\hline
\multirow{2}{*}{SPP low-$Q^{2}$} & Bias & $14.1\%$ & $18.6\%$ & $-6.11\%$ \\
                                & Size  & $2.0$\%  & $-1.6$\%  & $-2.2$\% \\
\hline\hline
\end{tabular}
\caption{Differences in the oscillation parameter constraints observed in the new and most impactful SDS. ``Bias'' shows changes to the center of the 2$\sigma$ confidence interval divided by the 1$\sigma$ Asimov interval width, and ``size'' is the change in size of the one-dimensional, 2$\sigma$, FC-corrected interval.}
\label{Tab:FakeData}
\end{table}

Various SDS were described in Ref.~\cite{T2K:2023smv}, and have also been carried out in this analysis. New SDS for this analysis are: replacing the default single-pion production model with the Martini et al. 1$\pi$ model~\cite{Martini:2009uj,Martini:2014dqa} to test the robustness of the model for the $\nu_{\mu}$CC1$\pi^+$-like SK sample; altering the default nuclear response to the Hartree-Fock Continuum Random Phase Approximation (HF CRPA)~\cite{Jachowicz:2002rr,Pandey:2014tza} to test the new proton-tagged ND280 samples. In total 19 SDS were performed. New SDS and those with the most significant impact are shown in \autoref{Tab:FakeData}. The observed bias on the center of the \dmsqtwothree 2$\sigma$ interval exceeded the bias condition for both the HF CRPA and non-CCQE SDS. As a result, the \dmsqtwothree contour was smeared by $3.1\times10^{-5}\;\text{eV}^2$, determined from the quadrature sum of the biases on \dmsqtwothree from all SDS. Additionally, a single SDS (SPP low-$Q^{2}$~\cite{MINERvA:2019kfr}) was found to change the $90\%$ confidence interval of \deltacp such that $\deltacp = \pi$ is not excluded.

\begin{figure}[htbp]
  \centering
  \includegraphics[width=0.48\textwidth]{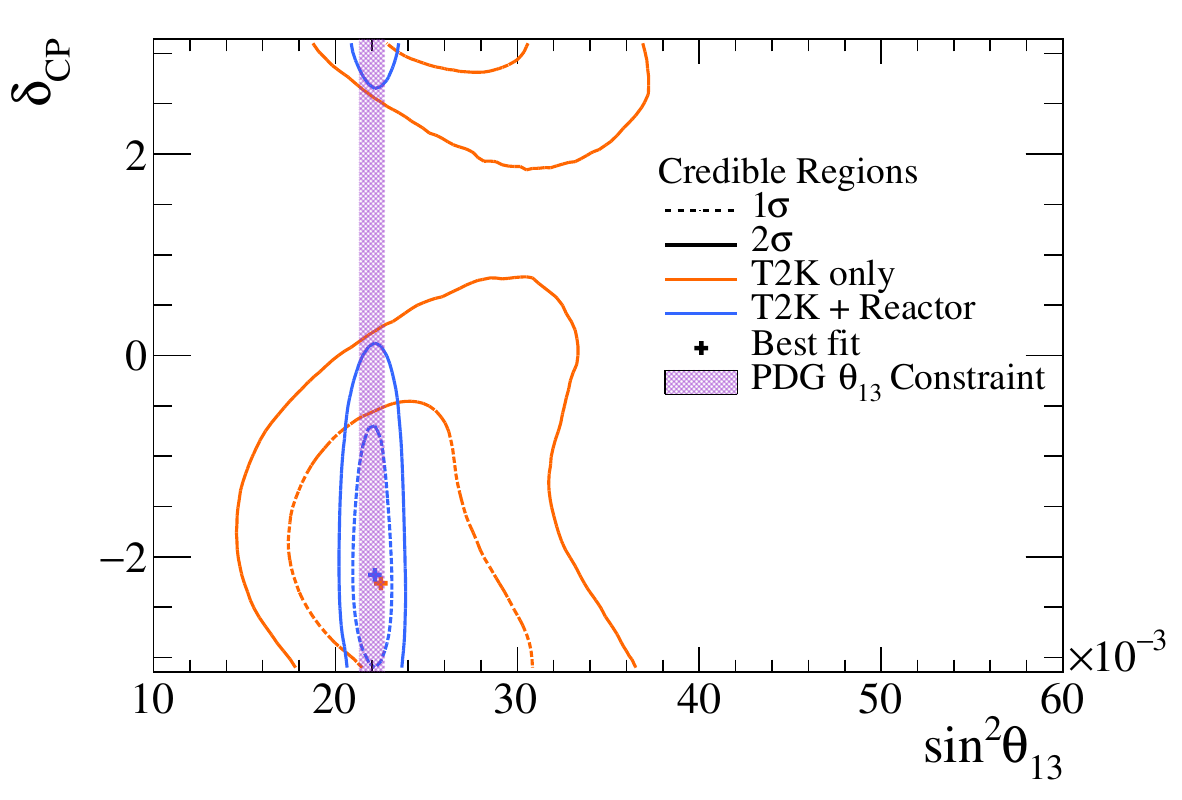}
  \caption{Credible regions in the \ssqthonethree--\deltacp plane produced with the MCMC analysis, shown with and without the RC applied, and overlaid with the RC constraint from Ref.~\cite{PDG2021}, for normal ordering.}
  \label{Fig::Comp_wRC}
\end{figure}
\indent{{\bf {\em Oscillation results and discussion---}}} \autoref{Fig::Comp_wRC} shows credible regions in the \ssqthonethree--\deltacp plane produced with and without the RC applied. The measurement of \ssqthonethree without the RC applied is consistent with the PDG value. When marginalized over both mass orderings, the best-fit value of \ssqthonethree with (without) the RC applied is $22.1\substack{+0.6 \\ -0.7} \times 10^{-3}$ ($23.5\substack{+5.6 \\ -3.1}\times 10^{-3}$).

\begin{figure}[htbp]
\centering
 \begin{subfigure}[t]{0.48\textwidth}
 \includegraphics[width=\textwidth]{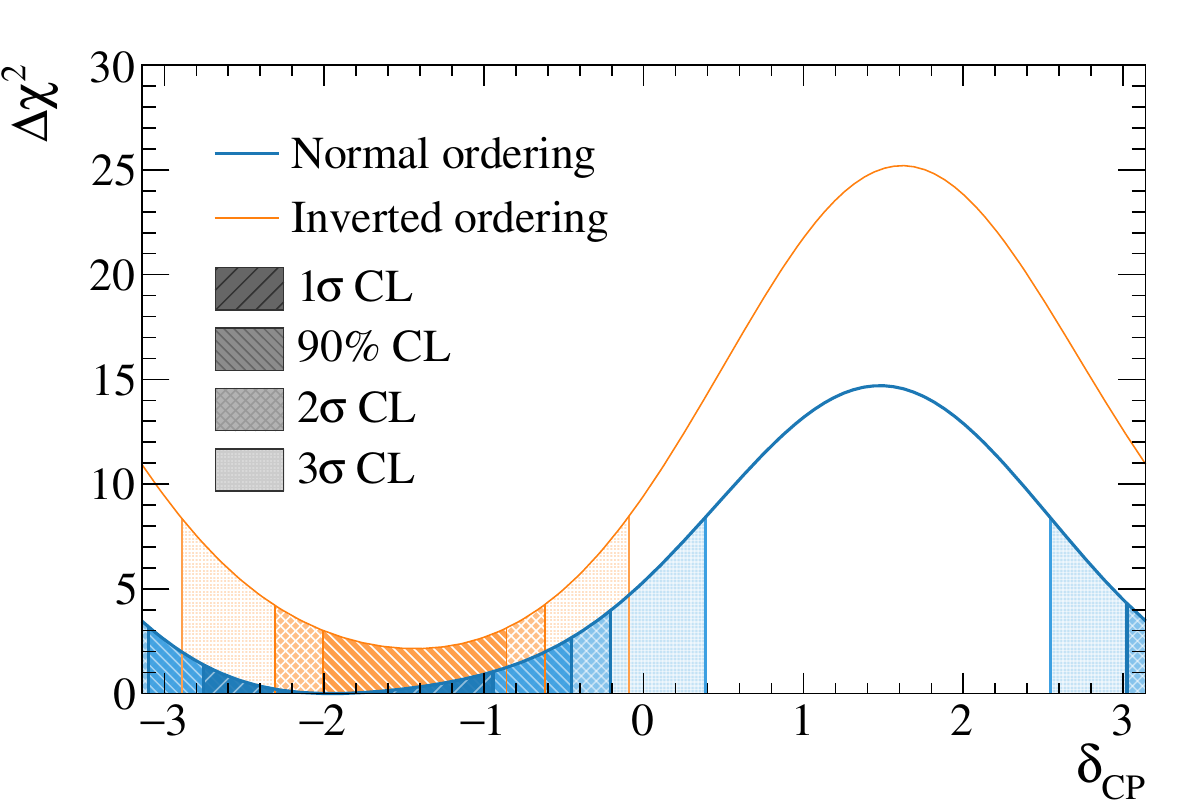}
 \end{subfigure}
\caption{The change in the best fit $\chi^{2}$ observed in the frequentist analysis as a function of \deltacp and the mass ordering. Shaded regions correspond to the FC-corrected confidence intervals.}
\label{Fig::contour_dcp_1D_wRC_FC}
\end{figure}

The FC-corrected frequentist confidence intervals for \deltacp are shown in \autoref{Fig::contour_dcp_1D_wRC_FC}. The best-fit value is $\deltacp=-2.18\substack{+1.22 \\ -0.47}$ (\deltacp$=-1.37\substack{+0.41 \\ -1.28}$) for normal (inverted) ordering with the RC applied. The data prefer values of \deltacp close to $-\frac{\pi}{2}$ radians, excluding values around $+\frac{\pi}{2}$ radians at $>$3$\sigma$ in both orderings. The CP-conserving values of 0 and $\pi$ are excluded at the 3$\sigma$ level in inverted ordering. In normal ordering, $\deltacp = 0$ is excluded at 90\% confidence, and although the nominal fit excludes $\deltacp=\pi$ at 90\% (\autoref{Fig::contour_dcp_1D_wRC_FC}), an SDS was found which could move the interval boundary past $\pi$, so this value is not excluded in the reported result. The result was statistically limited and can be expected to improve as more data is accumulated.

\begin{figure}[htbp]
\centering
 \begin{subfigure}[t]{0.48\textwidth}
 \includegraphics[trim={0 0 160pt 0}, clip, width=\textwidth]{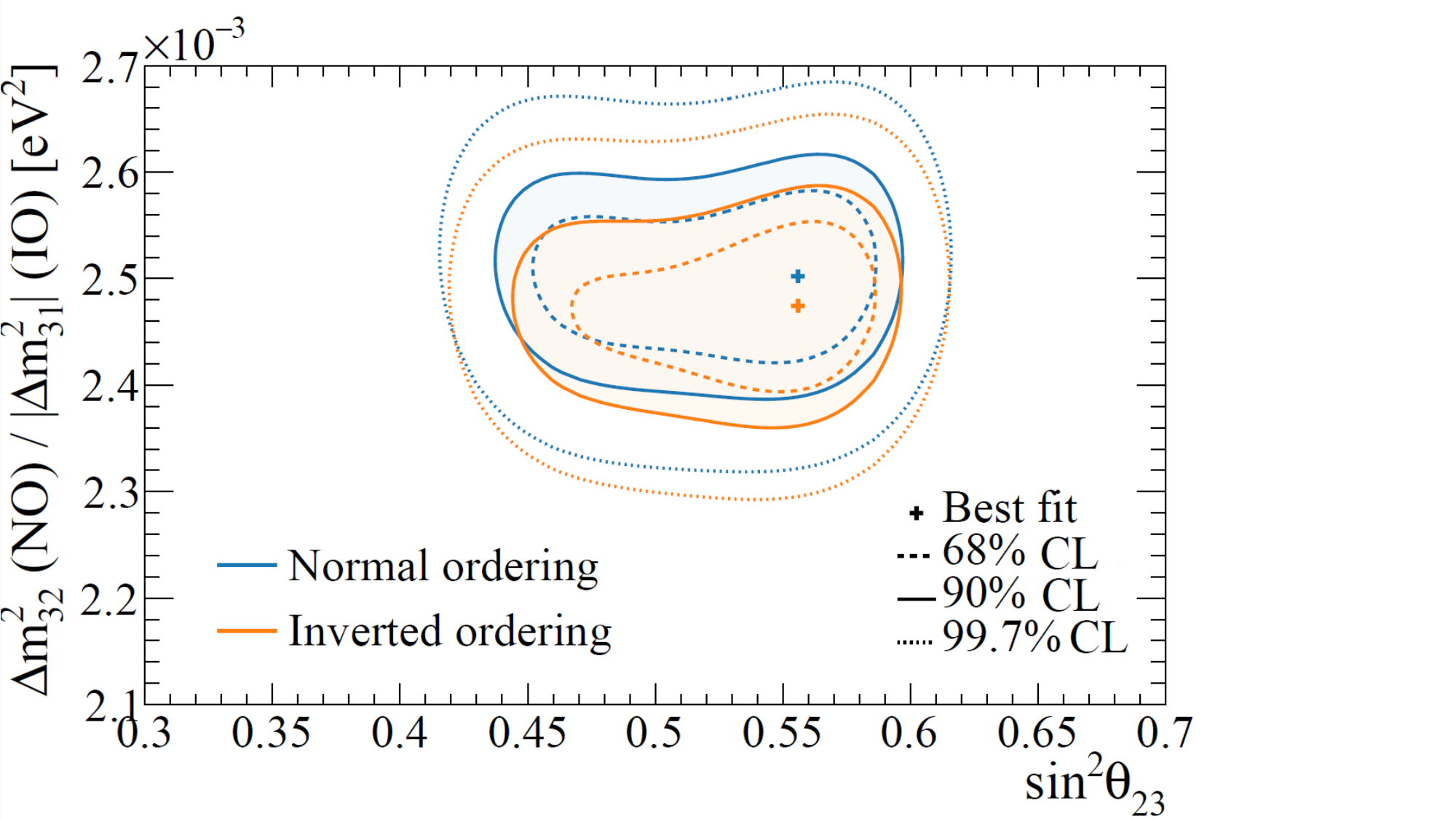}
 \end{subfigure}
\caption{Frequentist confidence intervals are shown in the \ssqthtwothree--\dmsqtwothree plane, produced using the constant $\Delta\chi^{2}$ method with the RC applied. The \dmsqtwothree contour is smeared to cover the SDS as described in the main text.}
\label{Fig::contour_dm2_32_th23_wRC}
\end{figure}

\autoref{Fig::contour_dm2_32_th23_wRC} shows frequentist confidence intervals in the \ssqthtwothree--\dmsqtwothree plane for both mass orderings. The data exhibit a weak preference for the upper octant of \ssqthtwothree and the normal ordering, with best-fit values of $\ssqthtwothree=0.559\substack{+0.018 \\ -0.078}$ and \dmsqtwothree$=(2.506\substack{+0.039 \\ -0.052})\times 10^{-3}$~\si{eV^{2}}. The MCMC analysis obtains a Bayes factor with (without) the RC applied of 2.3 (1.4) for the upper octant of $\theta_{23}$ over the lower and 2.7 (1.7) for the normal over inverted ordering. 

\begin{figure}[htbp]
\centering
 \begin{subfigure}[t]{0.48\textwidth}
 \includegraphics[page=1, width=\textwidth]{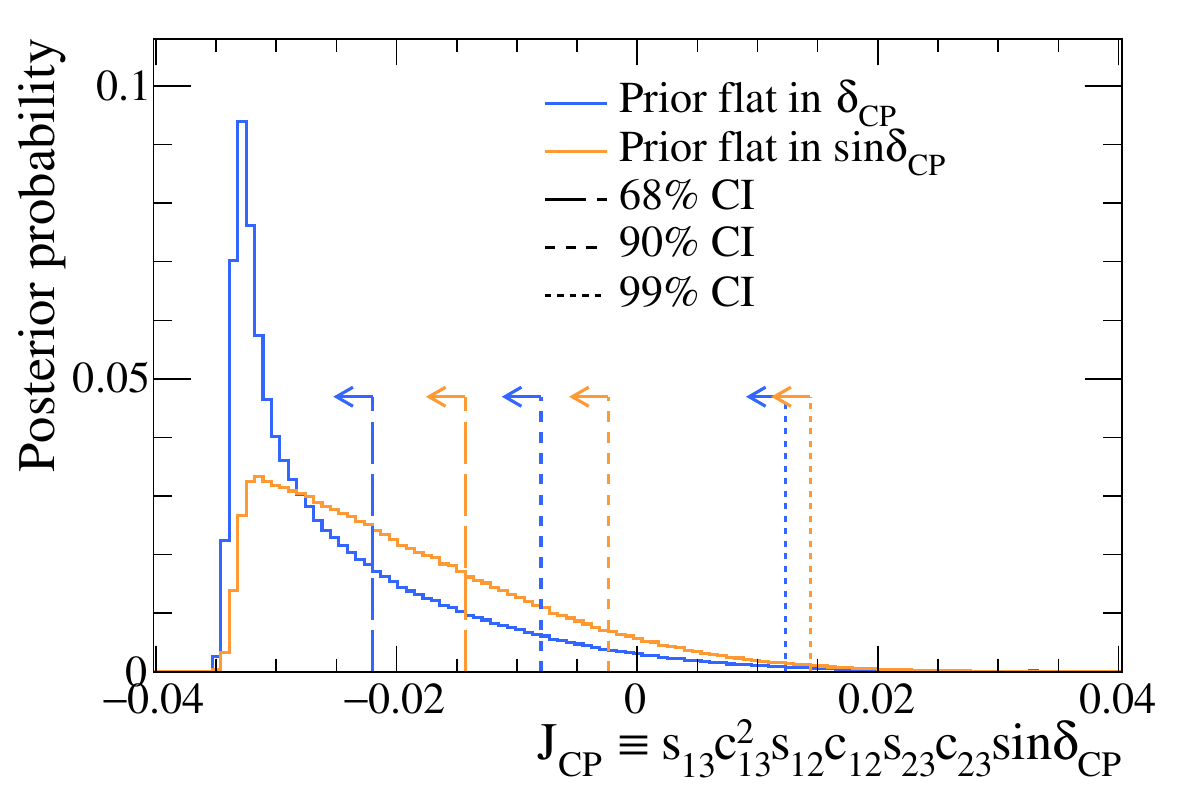}
 \end{subfigure}
\caption{Posterior probability distributions for the Jarlskog invariant taken from posterior distributions with priors that are either flat in \deltacp (blue) or flat in $\sin{\deltacp}$ (orange), obtained with the MCMC analysis for both orderings, with credible intervals (CIs) shown.}
\label{Fig::Jarlskog}
\end{figure}

The measurements presented thus far assume the PDG parameterization of the PMNS matrix. The Jarlskog invariant, \jarlcp, is a parameterization-independent way of measuring the scale of CP-violation generated by PMNS oscillations \cite{PhysRevLett.55.1039, Krastev:1988yu}. A zero (non-zero) value for \jarlcp indicates CP-conservation (CP-violation) in three-flavor neutrino mixing. The constraint on \jarlcp, obtained with the MCMC analysis, and the impact of the choice of \deltacp prior on that constraint, are shown in \autoref{Fig::Jarlskog}. The CP-conserving value, $\jarlcp = 0$, is excluded at the 90\% credible interval for both \deltacp priors: flat in \deltacp and flat in $\sin\deltacp$. Although changes in the prior were checked, the robustness of these credible intervals in \jarlcp has not been checked with SDS in this analysis.

\indent{{\bf {\em Conclusions---}}} The T2K collaboration has measured the three-flavor PMNS neutrino oscillation parameters \dmsqtwothree, \ssqthonethree, \ssqthtwothree, \deltacp, \jarlcp, and the mass ordering, using $3.6 \times 10^{21}$ POT at SK. The analysis includes a new $\nu_\mu$CC1$\pi^+$-like SK sample; new ND event samples in $\nu$-mode; significant improvements to the flux and neutrino interaction modeling; and a significantly expanded set of SDS to test the robustness of the analysis to out of model changes.
These improvements represent a significant step forward in the development of analysis methods which are robust to mis-modeling issues, a major obstacle for all current and future long-baseline neutrino oscillation experiments.
Our results show a weak preference for normal ordering and the upper octant of \ssqthtwothree with best-fit values of $\ssqthtwothree=0.559\substack{+0.018 \\ -0.078}$ and \dmsqtwothree$=(2.506\substack{+0.039 \\ -0.052})\times 10^{-3}$~\si{eV^{2}}. One of the CP conserving values, $\deltacp = 0$, is excluded at 90\% confidence, with a best-fit value of $\deltacp=-2.18\substack{+1.22 \\ -0.47}$ in normal ordering with the RC applied.

A data release including MCMC chains from the fit without the RC applied have been released in Ref.~\cite{this_zenodo}.

\indent{{\bf {\em Acknowledgements---}}}
The T2K collaboration would like to thank the J-PARC staff for superb accelerator performance. We thank the CERN NA61/SHINE Collaboration for providing valuable particle production data. We acknowledge the support of MEXT, JSPS KAKENHI  and bilateral programs, Japan; NSERC, the NRC, and CFI, Canada; the CEA and CNRS/IN2P3, France; the Deutsche Forschungsgemeinschaft (DFG 397763730, 517206441), Germany; the NKFIH  (NKFIH 137812 and TKP2021-NKTA-64), Hungary; the INFN, Italy; the Ministry of Science and Higher Education (2023/WK/04) and the National Science Centre (UMO-2018/30/E/ST2/00441, UMO-2022/46/E/ST2/00336 and UMO-2021/43/D/ST2/01504), Poland; the RSF (RSF 24-12-00271) and the Ministry of Science and Higher Education, Russia; MICINN  (PID2022-136297NB-I00 /AEI/10.13039/501100011033/ FEDER, UE, PID2024-157541NB-I00 (UAM) and PID2023-146401NB-I00 (US), Severo Ochoa Centres of Excellence Programme 2025-2029 (CEX2024001442-S)),  Government of Andalucia (FQM160) and the University of Tokyo ICRR's Inter-University Research Program FY2025 Ref. J1, and ERDF and European Union (UAM: H2020-MSCA-RISE-GA872549- SK2HK) and NextGenerationEU funds (PRTR-C17.I1) and Generalitat de Catalunya (AGAUR 2021-SGR-01506, CERCA program), University of Seville grant (RYC2022-035203-I funded by MICIU/AEI/10.13039/501100011033, ``ERDF a way of making Europe'' and FSE+, Ayudas ``Atracci\'{o}n de Investigadores con Alto Potencial'' Ref. VIIPPIT-2025), and Secretariat for Universities and Research of the Ministry of Business and Knowledge of the Government of Catalonia and the European Social Fund (2022FI\_B 00336), Spain; the SNSF and SERI, Switzerland; the STFC and UKRI, UK; the DOE, USA; and NAFOSTED (103.99-2023.144,IZVSZ2.203433), Vietnam. We also thank CERN for the UA1/NOMAD magnet, DESY for the HERA-B magnet mover system, the BC DRI Group, Prairie DRI Group, ACENET, SciNet, and CalculQuebec consortia in the Digital Research Alliance of Canada, and GridPP in the United Kingdom, the CNRS/IN2P3 Computing Center in France and NERSC, USA. In addition, the participation of individual researchers and institutions has been further supported by funds from the ERC (FP7), ``la Caixa'' Foundation, the European Union's Horizon 2020 Research and Innovation Programme under the Marie Sklodowska-Curie grant, the Horizon Europe Marie Sklodowska-Curie Staff Exchange project JENNIFER3 grant 101183137; the JSPS, Japan; the Royal Society, UK; French ANR and Sorbonne Universit\'{e} Emergences programmes; the VAST-JSPS (No. QTJP01.02/20-22); and the DOE Early Career program, USA. 
\bibliography{apssamp}

\end{document}